\documentclass[aps,prl,twocolumn,superscriptaddress]{revtex4}%
\usepackage{graphicx}
\usepackage{amsmath}
\usepackage{amsfonts}
\usepackage{color}
\usepackage{amssymb}%
\providecommand{\U}[1]{\protect\rule{.1in}{.1in}}

\begin{document}
\title{Spin fluctuations in Sr$_{2}$RuO$_{4}$ from polarized neutron scattering:
implications for superconductivity}
\author{P. Steffens}
\affiliation{$I\hspace{-.1em}I$. Physikalisches Institut, Universit\"at zu K\"oln,
Z\"ulpicher Str. 77, D-50937 K\"oln, Germany}
\affiliation{Institut Laue Langevin,71 avenue des Martyrs, 38000 Grenoble, France}
\author{Y. Sidis}
\affiliation{Laboratoire L\'eon Brillouin, C.E.A./C.N.R.S., F-91191 Gif-sur-Yvette CEDEX, France}
\author{J. Kulda}
\affiliation{Institut Laue Langevin,71 avenue des Martyrs, 38000 Grenoble, France}
\author{Z. Q. Mao}
\affiliation{Department of Physics, Graduate School of Science, Kyoto University, Kyoto
606-8502, Japan}
\affiliation{Department of Physics, Tulane University, New Orleans, LA 70118, USA}
\author{Y. Maeno}
\affiliation{Department of Physics, Graduate School of Science, Kyoto University, Kyoto
606-8502, Japan}
\author{I.I. Mazin}
\affiliation{Code 6393, Naval Research Laboratory, Washington, District of Columbia 20375, USA}
\author{M. Braden}
\email{[e-mail: ]braden@ph2.uni-koeln.de}
\affiliation{$I\hspace{-.1em}I$. Physikalisches Institut, Universit\"at zu K\"oln,
Z\"ulpicher Str. 77, D-50937 K\"oln, Germany}

\begin{abstract}
Triplet pairing in Sr$_{2}$RuO$_{4}$ was initially suggested based on the
hypothesis of strong ferromagnetic spin fluctuations. Using polarized
inelastic neutron scattering, we accurately determine the full spectrum of
spin fluctuations in Sr$_{2}$RuO$_{4}$. Besides the well-studied
incommensurate magnetic fluctuations we do find a sizeable quasiferromagnetic
signal, quantitatively consistent with all macroscopic and microscopic probes.
We use this result to address the possibility of magnetically-driven triplet
superconductivity in Sr$_{2}$RuO$_{4}$. We conclude that, even though the
quasiferromagnetic signal is stronger and sharper than previously anticipated,
spin fluctuations alone are not enough to generate a triplet state
strengthening the need for additional interactions or an alternative pairing scenario.

\end{abstract}

\pacs{7*******}
\maketitle












Superconducting Sr$_{2}$RuO$_{4}$ \cite{1,2,3} was proposed to be a
solid-state analogue of He$^{3},$ $i.$ $e.,$ a triplet
superconductor \cite{Rice,mazin97a}, based on its proximity to SrRuO$_{3},$ a
ferromagnetic (FM) metal. A simple model derived from the density-functional
theory (DFT) for SrRuO$_{3},$ CaRuO$_{3}$ and SrYRu$_{2}$O$_{6}$
\cite{MazinRu} ascribed the mass and spin susceptibility
renormalization to FM fluctuations, and predicted a triplet
pairing \cite{mazin97a}. Experimental evidence pointing toward a particular
(chiral-$p$) triplet was obtained, such as temperature-independent uniform
susceptibility for the in-plane fields and time-reversal symmetry
breaking \cite{2,kallin-a,5,npjQM}. However, the dominant spin fluctuations in
Sr$_{2}$RuO$_{4}$ are not FM (\textit{i.e.} $q$=0), but incommensurate (IC)
antiferromagnetic (AFM) \cite{mazin99a,sidis99a}, and several experiments are
inconsistent with either triplet states, or time-reversal breaking, or both
\cite{npjQM}. Various theories were proposed to explain triplet pairing by
incorporating higher-order vertex corrections \cite{nomura00,nomura02}, the
interplay of incommensurate charge and spin fluctuations \cite{raghu10} or
orbital fluctuations \cite{takimoto00,tsuchiizu15}, arriving at different
superconducting states. Even the question about which bands drive pairing
remains controversial \cite{huo13,scaffidi14}.

The Fermi surface of Sr$_{2}$RuO$_{4}$ is known to tiny
details \cite{2,bergemann03a,veenstra14,kim18}. It has two
quasi-one-dimensional (q1D), and one rather isotropic quasi-two-dimensional
(q2D) sheets, derived from $d_{xz,yz}$ and $d_{xy}$ orbitals, respectively.
Sr$_{2}$RuO$_{4}$ exhibits an\ almost temperature independent normal-state
susceptibility \cite{maeno97}, which is enhanced by a factor $\sim$7 compared
to the DFT value \cite{oguchi95,singh95a,hase96}. The enhancement factor of
the IC fluctuations is even larger, $\sim$30 \cite{sidis99a,braden02b,servant}%
, since the bare susceptibility is larger \cite{mazin99a}. Also the electronic
specific heat coefficient of about 38~mJ/mol$\cdot$K$^{2}$ is enhanced by a
factor of $\sim$3, yielding a Wilson ratio of $\sim$2. Similarly, quantum
oscillations show strong and band-dependent mass renormalizations, which can
be explained by quasiferromagnetic (qFM) fluctuations \cite{mazin97a}, in the
spirit of He$^{3}$, but also in terms of local Hund's rule
fluctuations \cite{Licht,mravlje}.

Inelastic neutron scattering (INS)  experiments detect strong IC spin
fluctuations at $\mathbf{q}_{\mathrm{IC}}$=($\pm$0.3,$\pm$0.3,$q_{l}%
$) \cite{sidis99a,braden02b,servant,braden2004a,iida,iida2,sro-gap} arising
from nesting in the q1D bands. Upon minor substitution with Ti or Ca this
instability condenses into a static spin-density wave with the same
$\mathbf{Q}$ \cite{21,22,23,24}. INS also assesses the anisotropy of magnetic
excitations, which is known to favor triplet
pairing \cite{ogata,ogata1,ogata2}, and find it to be non-negligible, but still
small \cite{braden2004a}. Finally, recent high-resolution INS reveals that the
nesting fluctuations do not change between the normal and superconducing
states even for energies well below the superconducting gap \cite{sro-gap}.
The NMR relaxation rate, $1/T_{1}T$, probes the spin susceptibility
$\chi^{\prime\prime}(\mathbf{q,\omega})/\omega$ integrated over the entire
Brillouin zone, and exhibits the same temperature dependence as the INS
nesting signal \cite{sidis99a,braden02b,ishida01,ishida01a}, indicating that
it is dominated by the latter. However, $1/T_{1}T$ also shows a weaker,
temperature-independent offset, pointing to another contribution tentatively
attributed to the FM response. This tendency towards ferromagnetism can be
enhanced by Co \cite{ortmann13} or Ca \cite{nakatsuji03} substitution.

To this end, we have used polarized INS to search for the missing FM
fluctuations in Sr$_{2}$RuO$_{4}$. The magnetic response consists of
two components: a broad maximum around ${\mathbf{q}}=0,$ which we will call
qFM, and an IC, and much stronger, AFM component. We entered this full
magnetic susceptibility into the BCS equations describing
spin-fluctuation-induced superconducting pairing.

Because neutron polarization analysis suffers from a reduced intensity, we used a large sample of ten
aligned crystals grown at Kyoto University \cite{mao00} with a total volume of 2.2\ cm$^{3}$ and a mosaic spread of
1.9(2) degrees. 
Experiments were
performed on the spectrometer IN20 at the Institut Laue Langevin
\cite{exp-det} In general, neutron scattering only senses magnetic components
that are polarized perpendicular to the scattering vector $\mathbf{Q}$. The
polarization analysis distinguishes spin-flip (SF$_{i}$ with $i$=$x$, $y$, and
$z$ the direction of neutron polarization) and non-spin-flip (nSF$_{i}$)
processes and adds further selection rules. Phonon scattering and nuclear
Bragg peaks only contribute to the nSF$_{i}$ channels, but magnetic scattering
contributes to the SF$_{i}$ channel when the magnetic component is
perpendicular to the direction of neutron polarization, and to the nSF$_{i}$
channel otherwise. We use the conventional coordinate system with $\mathbf{x}$
parallel to $\mathbf{Q}$, $\mathbf{z}$ perpendicular to the
scattering plane, and $\mathbf{y}=\mathbf{z}\times\mathbf{x}$.

\begin{figure}[ptb]
\includegraphics[width=1.3\columnwidth]{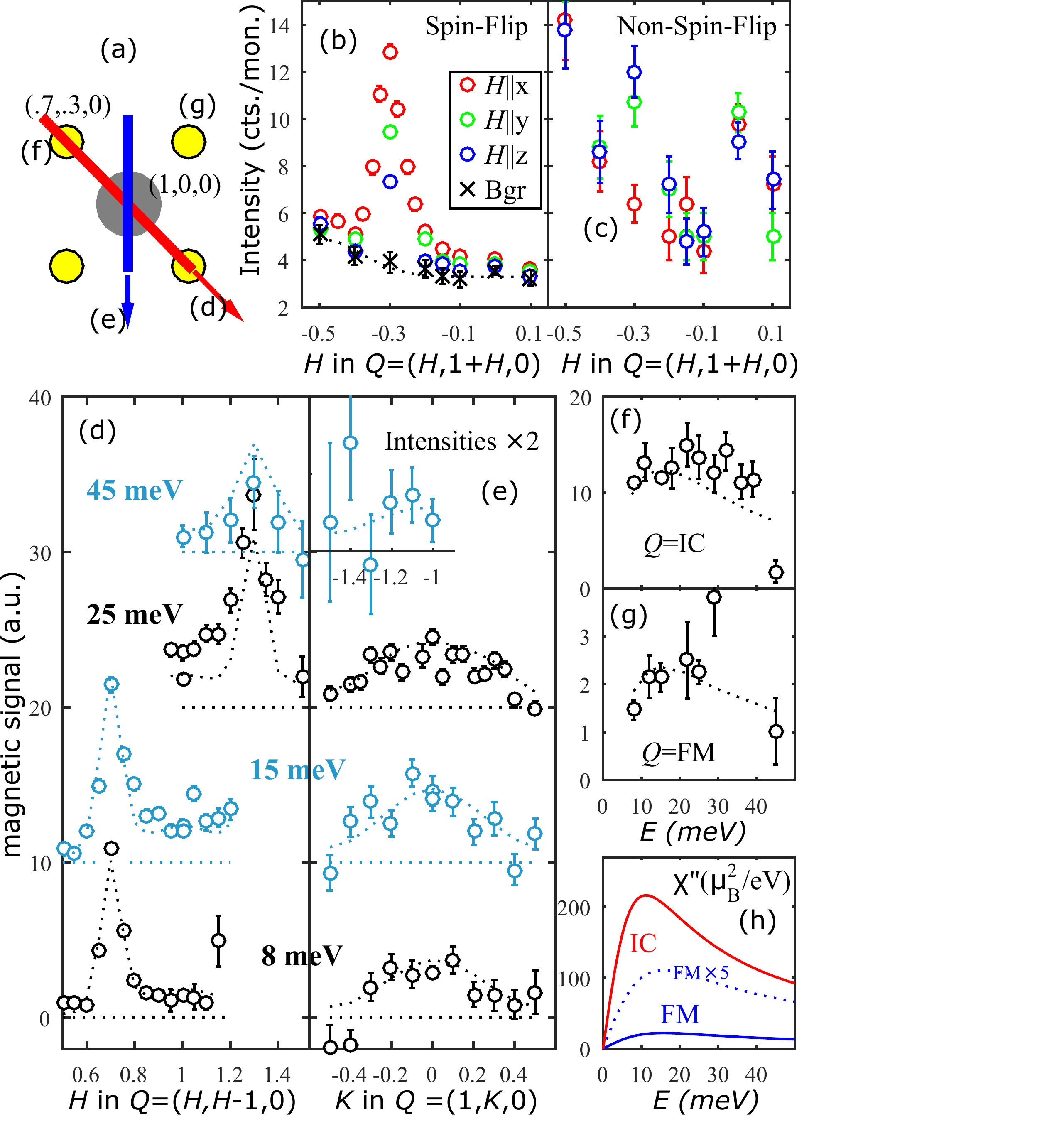}
\caption{(a) 2D reciprocal space of Sr$_{2}$RuO$_{4}$; qFM
scattering is indicated by large (grey) discs and the IC signal by small
(yellow) circles. Arrows show typical scan directions. (b-c) Diagonal
scans at 8~meV and 1.6~K (across (-0.3,0.7,0) and (0,1,0)): (b) SF count
rates, (c) nSF count rates. (d) Magnetic signal along diagonal scans at 1.6~K;
note that the BG is eliminated through the polarization analysis. Scan paths
are not identical, but all run through one $\mathbf{Q}_{\mathrm{IC}}$ towards
(1,0,0), see (a). The signal has been corrected for the magnetic form factor
and the Bose factor and represents $\chi^{\prime\prime}(\mathbf{q}, E)$
convoluted with the resolution function. In (e) the results of the scans
parallel to the a*/b* axes are shown. Energy scans at $\mathbf{Q}%
_{\mathrm{IC}}$ \ and $\mathbf{Q}_{\mathrm{FM}}$ \ are shown in (g) and (g),
respectively. The corresponding susceptibility is shown in absolute units in
(h). Lines in (d-g) denote the fitted model. }%
\label{sro-pola2}%
\end{figure}

Even with our large sample it was impossible to quantitatively analyze the qFM
response by unpolarized INS, because it is too little structured in
$\mathbf{q}$ space impeding a background (BG) determination, see supplemental
material \cite{suppl}. In contrast, the polarization analysis permits a direct
BG subtraction at each point in $\mathbf{Q}$ and energy. For instance,
2I(SF$_{x}$)-I(SF$_{y}$)-I(SF$_{z}$) yields a BG-free total magnetic signal
(up to a correction for the finite flipping ratio). Fig. \ref{sro-pola2} (b-c)
shows a representative scan through both the IC and the FM $\mathbf{Q}$
positions. The full polarization analysis is shown for the SF (b) and the nSF
(c) channels. The SF signals have been counted with better statistics, because
the SF count rates always contain the magnetic signal and have a lower BG.
Only the nSF$_{y}$ and nSF$_{z}$ channels contain a single magnetic component
superposed with the larger nSF scattering, which contains all the phonon
contributions. The appearance of the nesting signal in various channels is
well confirmed; Fig. \ref{sro-pola2} (b) clearly shows the anisotropy of the
IC nesting signal at $(-0.3,0.7,0)$ discussed in Ref. \cite{braden2004a}. The
sharp enhancement at $(0,1,0)$ is present only in the nSF channel, which
proves its non-magnetic character (the longitudinal zone-boundary phonon)
\cite{braden-phon1,braden-phon2}. The finite flipping ratio was determined on
several phonon modes, which integrates the signal of all individual crystals, 
yielding values between 8 and 10. The final analysis only used the
SF data, corrected by the average flipping ratio, because of their higher signal to BG ratio \cite{noteBG}.

Polarized INS results displaying the sum of the two magnetic components
(in-plane plus out-of-plane) are shown in Fig. 1 for $T$=1.6~K and in the
supplemental material for $T$=150~K \cite{suppl}. In order to compare scans
taken at different but equivalent scattering vectors, a correction for the
magnetic form factor has been applied. The observation of magnetic
fluctuations in so many different scans unambiguously documents the existence
of sizeable qFM fluctuations. The analysis furthermore yields the absolute
scale of the magnetic response throughout the entire Brillouin zone, which
allows us to construct a model for the full susceptibility $\chi^{\prime
\prime}(\mathbf{q},E)$. The calibration into
absolute susceptibility units has been performed by the comparison with the
scattering intensity arising from an acoustic phonon, similar to the procedure
described in Ref. \onlinecite{qureshi2014}. This calibration can be performed
with high precision in the case of Sr$_{2}$RuO$_{4}$, because the phonon
dispersion is well known and a lattice dynamical model exists that was used to
calculate the phonon signal strength at finite propagation vectors
\cite{braden-phon1,braden-phon2}, while in most cases the $q\rightarrow0$
limit is used as an approximation.

The quantitative model fitted to the data consists of two parts: the IC peaks
centered at $\mathbf{Q}_{\mathrm{IC}}$ and the broad and weakly $\mathbf{q}%
$-dependent qFM part at the zone center. We write $\chi^{\prime\prime
}(\mathbf{q},E)=\chi_{\mathrm{IC}}^{\prime\prime}(\mathbf{q},E)+\chi
_{\mathrm{FM}}^{\prime\prime}(\mathbf{q},E)$, where
\begin{equation}%
\chi_{\mathrm{IC}}^{\prime\prime}({\mathbf{q},E})=\chi_{\mathrm{IC}}^{\prime
}\frac{\Gamma_{\mathrm{IC}}\cdot E}{E^{2}+\Gamma_{\mathrm{IC}}^{2}%
[1+\xi_{\mathrm{IC}}^{2}(\frac{2\pi}{a}\Delta{q})^{2}]^{2}}\text{~ ~ ~}%
\end{equation}
is the single-relaxor formula with both $(\Gamma_{\mathbf{q}})^{-1}$ and
$\chi^{\prime}(\mathbf{q},0)$ decaying with the same correlation length
$\xi_{\mathrm{IC}}.$ Here $\Delta{q}=|\mathbf{q}-\mathbf{q}_{\mathrm{IC}}|$,
and is measured in the reciprocal lattice units, (r.l.u. ), equal to $2\pi/a$.

Equation (1) describes a typical magnetic response near an AFM instability
\cite{buchmoriya}.
The qFM term was described by a broad Gaussian, and its energy dependence in
the single-relaxor form with the constant parameter $\Gamma_{\mathrm{FM}}$:
\begin{equation}
\chi_{\mathrm{FM}}^{\prime\prime}(q,E)=\chi_{\mathrm{FM}}^{\prime}\cdot
\frac{\Gamma_{\mathrm{FM}}\cdot E}{E^{2}+\Gamma_{\mathrm{FM}}^{2}}%
\cdot\text{exp}\left(  -\frac{q^{2}}{W^{2}}4\ln(2)\right)  \label{sro-sus2}%
\end{equation}
and $\mathbf{q}$ is the distance to the nearest 2D Bragg point. The parameters
resulting from a global fit to the whole data set are given in Table I
\cite{alternativ}. The model susceptibility was convoluted with the
spectrometer resolution using the reslib program package \cite{reslib} and
scaled through phonon scattering \cite{braden-phon2} yielding the lines in Fig.
1 (d-h).

\begin{table}[ptb]%
\begin{tabular}
[c]{lcccccc}\hline
$T$ & $\chi^{\prime}_{\mathrm{FM}}$ & $W$ & $\Gamma_{\mathrm{FM}}$ &
$\chi^{\prime}_{\mathrm{IC}}$ & $\xi_{\mathrm{IC}}$ & $\Gamma_{\mathrm{IC}}$\\
$[$K$]$ & [$\mu_{B}^{2}/eV$] & [r.l.u] & [eV] & [$\mu_{B}^{2}/eV$] & [\AA ] &
[eV]\\\hline
1.6 & 22$\pm$1 & 0.53$\pm$0.04 & 15.5$\pm$1.4 & 213$\pm$10 & 9.7$\pm$0.5 &
11.1$\pm$0.8\\
150 & 22$\pm$2 & 0.47$\pm$0.06 & 19.0$\pm$3.5 & 89$\pm$7 & 6.1$\pm$0.5 &
17.8$\pm$2.9\\
&  &  &  &  &  & \\\hline
& qFM T & qFM S & IC T & IC S & total T & total S\\
$\chi^{\prime}$ & 10.6 & 0.21 & 16.8 & 94.8 & 18 & 87\\
$\chi^{\prime}_{zz}$ & 11.7 & 0.23 & 29.1 & 164.2 & 30.3 & 155.6\\
$\chi^{\prime}_{ab}$ & 9.6 & 0.19 & 9.7 & 54.7 & 11.3 & 48.3\\
&  &  &  &  &  &
\end{tabular}
\caption{(Upper part) Parameters of the $\chi^{\prime\prime}(\mathbf{q},E)$
model for Sr$_{2}$RuO$_{4}$ refined with the polarized INS data for $T$=1.6
and 150\,K. (Lower part) The largest triplet, T, and singlet, S, eigenvalues
(in arbitrary
units)
of the interaction matrices $V_{s}$ and $V_{t}$, respectively (Eq. 3),
obtained for the isotropic susceptibility, $\chi^{\prime}$=$\chi^{\prime
}(q,0)$ or for the anisotropic components $\chi^{\prime}_{zz}$ and
$\chi^{\prime}_{ab}$; the largest eigenvalues for qFM or IC fluctuations only
are shown together with those for the total susceptibility.}%
\end{table}

The corresponding real part of the susceptibility at zero energy $\chi
^{\prime}(\mathbf{q},E=0)$, the amplitudes of the spectra at fiexd $\mathbf{q}$, is displayed in Fig. 2. The qFM signal shows no
significant anisotropy and corresponds to the macroscopic susceptibility,
which also exhibits only weak anisotropy \cite{2,3}. For the IC peak, the model
describes the average of the in-plane and out-of-plane susceptibilities
\cite{braden2004a}, with $\chi_{c}^{\prime}$ ($\chi_{ab}^{\prime})$ slightly
larger (smaller) than this value. The model was obtained by refining the only
6 parameters with the total set of 120 independent data points at 1.6~K and 76
at 150~K. Thus obtained $\chi_{\mathrm{IC}}^{\prime}$ and $\Gamma
_{\mathrm{IC}}$ are somewhat higher than those extracted from unpolarized INS
\cite{sidis99a,braden02b}. The correlation length $\xi_{\mathrm{IC}}$ is less
accurate but the qualitative decrease at higher temperature is unambiguous. In
principle, one should consider the in-plane and out-of-plane components of the
IC peak separately and then take their superposition, but the limited
statistics does not allow for that. In contrast to the IC signal, the qFM one
is basically temperature-independent, in agreement with the macroscopic
measurement \cite{maeno97}. Thus, the qFM response becomes more visible at
high temperatures. Note that, due to the simplicity of the model
\cite{alternativ}, the  macroscopic susceptibility of $\sim$28\thinspace
$\mu_{B}^{2}/$eV$\cdot$(f.u.) is smaller than in the model, $\sim$%
41\thinspace$\mu_{B}^{2}/$eV$\cdot$(f.u.).

The model $\chi^{\prime\prime}(\mathbf{q},E)$ can also be successfully
verified against $1/{T_{1}T}$ in NMR
\cite{imai98a,ishida01,berthier96,ishida01a,mukuda98} and with specific heat
data \cite{maeno97,specheat1,specheat2}, see supplemental material
\cite{suppl}. The impact of the qFM fluctuations must not be underestimated;
because of the larger phase space, they yield about 85\% of the specific-heat enhancement.

The qFM response does not correspond to the paramagnon scattering expected
close to a FM transition \cite{buchmoriya}; instead it can be viewed as as an
AFM instability with a small but finite propagation vector near the
Brillouin-zone center and a width that largely exceeds the length of the
propagation vector. The superposition of several low-$\mathbf{q}$
contributions can result in the observed broad feature centered at
$\mathbf{q}$=(0,0). Ca$_{2-x}$Sr$_{2}$RuO$_{4}$ with $0.2<x<0.5$ shows a
cluster-glass FM ordering or a metamagnetic transition into a state with
sizeable polarized moment \cite{nakatsuji03}. Also in these nearly FM
compounds the magnetic fluctuations still differ from the FM paramagnon
response and retain a small-$q$ incommensurate AFM character
\cite{friedt04,steffens07,steffens11}, but the width in the Ca-substituted
materials is much smaller than that of the qFM part in Sr$_{2}$RuO$_{4}$. The qFM
signal furthermore exhibits a characteristic energy that is only a little
larger than that of the IC signal, supporting the notion that Sr$_{2}$%
RuO$_{4}$ is also close to FM order \cite{ortmann13,nakatsuji03}.

\begin{figure}[ptb]
\vskip -0.5 cm
 \includegraphics[width=0.7\columnwidth]{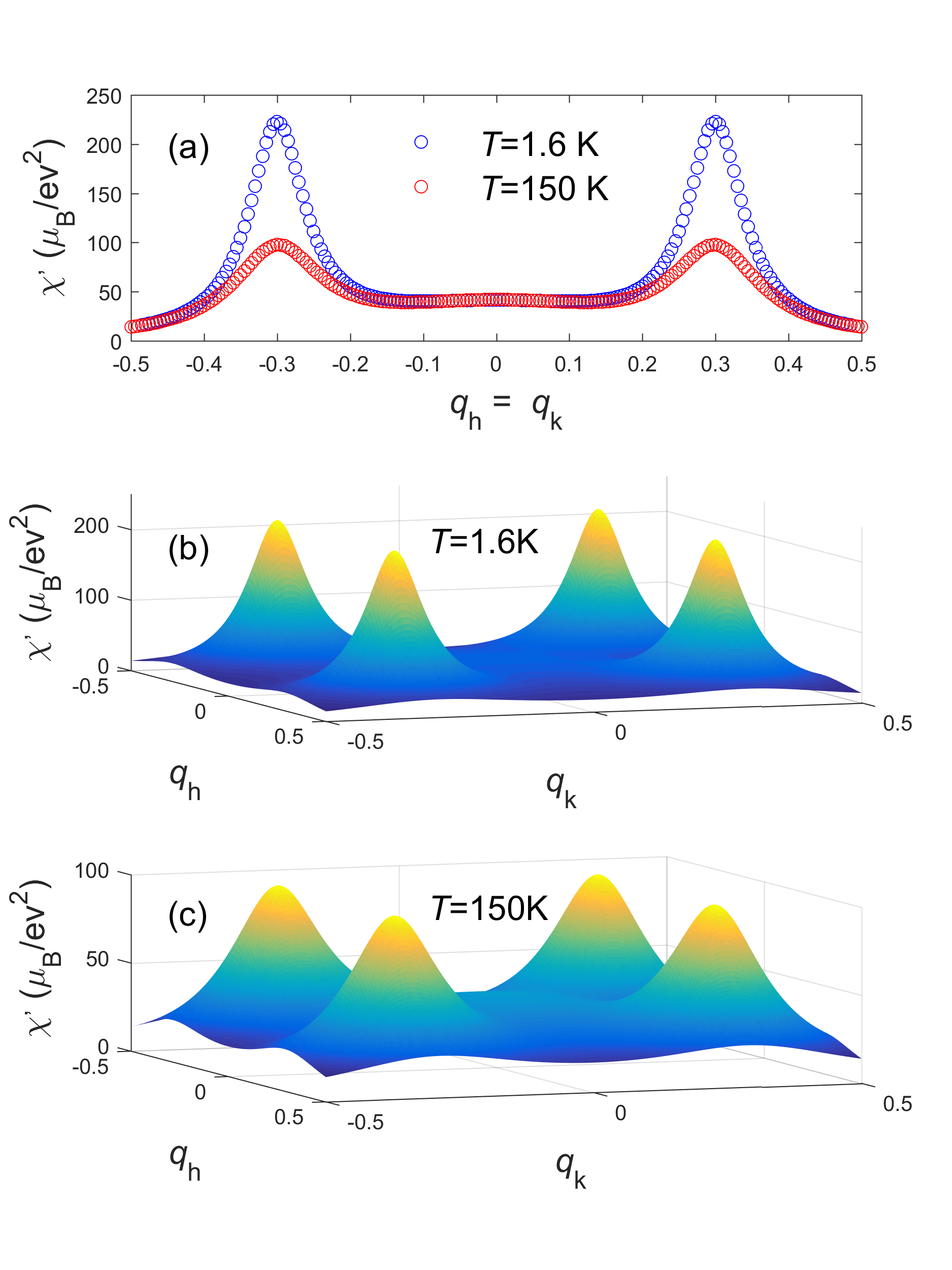}
\vskip -0.5 cm
\caption{ The real part of the static susceptibility
$\chi^{\prime}({\mathbf{q}},E=0)$ as described by eqs. (1,2)
 along the zone diagonal (a) and for the entire zone (b)
at 1.6\,K and (c) at 150\,K.}%
\label{sro-real1}%
\end{figure}

To access the role of the qFM fluctuations in the superconducting pairing we
have used the standard weak-coupling approach relating the spin-mediated
pairing interaction $V(\mathbf{k},\mathbf{k}^{\prime})$ to the full
$\chi(\mathbf{q},E)$, see, e.g. Ref. \cite{Eremin_review}: $\lambda
\mathbf{\Delta}(\mathbf{k})=\sum_{\mathbf{k}^{\prime}}V(\mathbf{k}%
,\mathbf{k}^{\prime})\cdot\mathbf{\Delta}(\mathbf{k}^{\prime})$, where
$\Delta(\mathbf{k})$ characterizes the superconducting order parameter (SOP).
$V(\mathbf{k},\mathbf{k}^{\prime})$ is, for the singlet and triplet
pairings \cite{Eremin_review}:%

\begin{align}
&  V_{s}(\mathbf{q=k}-\mathbf{k}^{\prime})=-3I^{2}(\mathbf{q)}\chi^{\prime
}(\mathbf{q},0)\frac{1}{\sqrt{v_{F}(\mathbf{k})v_{F}(\mathbf{k^{\prime}})}%
}\label{V}\\
&  V_{t}(\mathbf{q=k}-\mathbf{k}^{\prime})=I^{2}(\mathbf{q)}\chi^{\prime
}(\mathbf{q},0)\frac{\mathbf{\hat{v}}_{F}(\mathbf{k})\cdot\mathbf{\hat{v}}%
_{F}(\mathbf{k^{\prime}})}{\sqrt{v_{F}(\mathbf{k})v_{F}(\mathbf{k^{\prime}})}%
}\nonumber
\end{align}
where $I(\mathbf{q)}$ is defined as $I({q})=\chi_{0}(\mathbf{q})^{-1}%
-\chi(\mathbf{q})^{-1},$ and $\chi_{0}(\mathbf{q})$ is the noninteracting
(Lindhardt) susceptibility. Note that only the amplitude of the single-relaxor
spectra, $\chi^{\prime}(\mathbf{q},0)$, enters the interaction matrices. We
use the tight binding Hamiltonian of Ref. \cite{okamoto-millis} and
parameterize the interaction as \cite{mazin99a}: $I(\mathbf{q)}=I(0)/[1+b(\frac
{a}{\pi})^{2}q^{2}]$. The matrices $V_{s,t}(\mathbf{k},\mathbf{k}^{\prime})$
are diagonalized by discretizing the Fermi surface into 1301 vectors
$\mathbf{k}$. The largest eigenvalue defines the the most stable
superconducting state, and the corresponding eigenvector the symmetry and the
structure of the SOP. The interaction parameter $I(q)$ is crucial. Based on
their calculations for SrRuO$_{3},$ Mazin and
Singh \cite{mazin97a,mazin98,mazin99a} assigned the $q$ dependence of $I$ to
the Hund's rule coupling on oxygen, and estimate $b=0.08.$ In the experiment,
we find a much larger value $b=0.44$, see Fig. 3(a), thus favoring more the
triplet pairing.

\begin{figure}[ptb]
\vskip -0 cm \includegraphics[width=0.85\columnwidth]{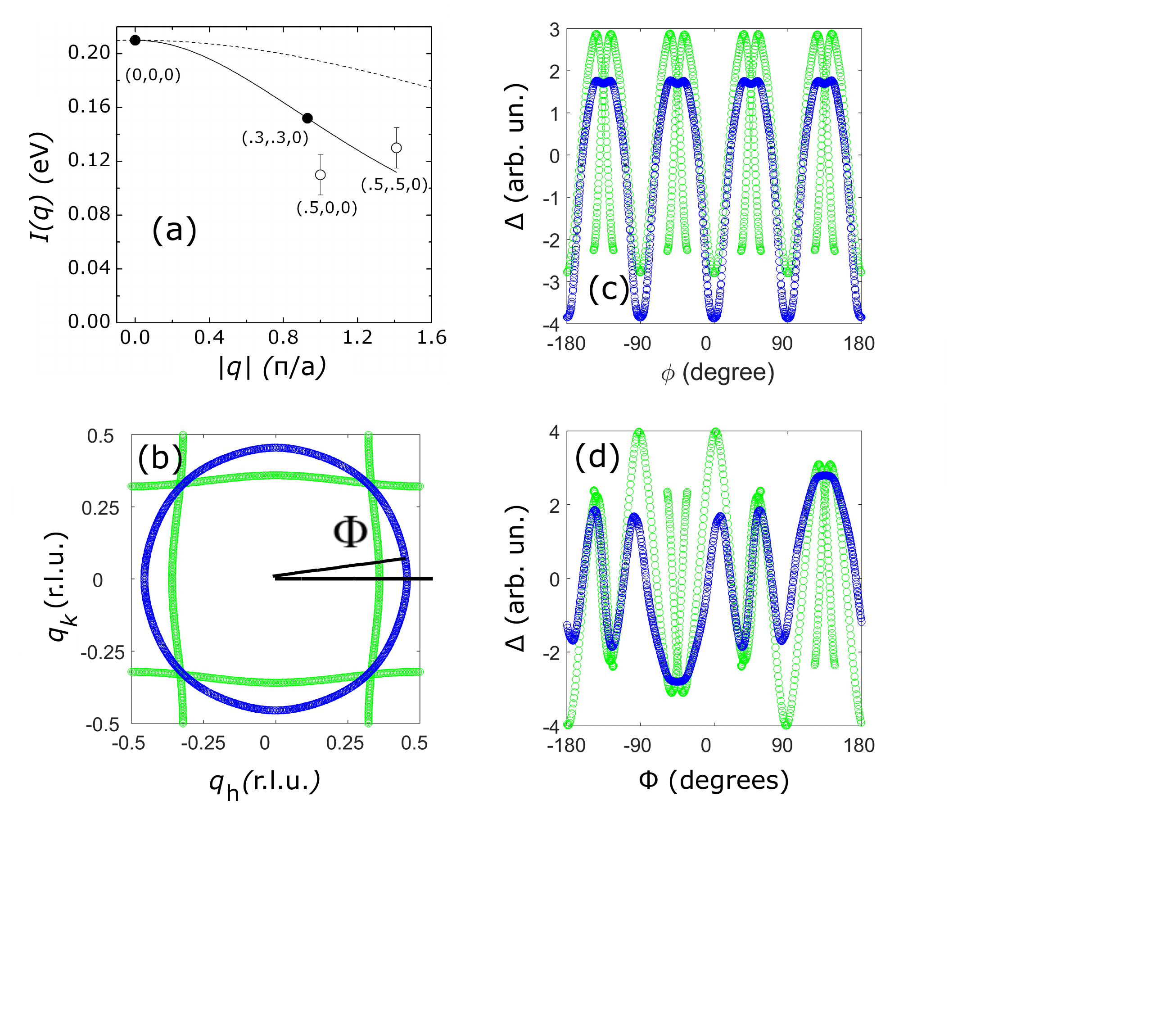} \vskip -1.2
cm\caption{ {(a) The interaction $I(\mathbf{q)}=I(0)/[1+b(\frac{a}{\pi}%
)^{2}q^{2}]$ for b=0.08 (dashed line) and b=0.44 (solid line) compared with
the experimental estimates; (b) 2D Fermi surface of Sr$_{2}$RuO$_{4}$ with
the q1D (green) and q2D (blue) sheets; (c,d) order parameter on
these sheets plotted against the angle with respect to the
$k_{x}$-axis for the most stable singlet (c) and triplet (d) solutions. } }%
\label{sro-calc}%
\end{figure}

To elucidate the role of the IC and qFM spin fluctuations, we have
diagonalized the matrices described by equation (3) using the two
contributions separately, and using the total $\chi^{\prime}=\chi^{\prime
}(\mathbf{q},0)$. The results are shown in Table I. As expected, for the IC
fluctuations alone singlet solutions are most stable, and the qFM ones give
triplets. With the total susceptibility, the IC fluctuations significantly
contribute to the triplet solution as well, but the graund state is still a
singlet: the ratio of the largest singlet to the largest triplet eigenvalue is
rather high, $R_{s/t}=$4.8 \cite{alternativ}. Even a five times larger qFM
part (clearly incompatible with the experiment) only reduces the ratio to
$R_{s/t}$=1.4. Sharpening the parameter $I(q)$ significantly helps the triplet
case, but not enough; tripling $b$ to 1.32 only reduces $R_{s/t}$ to 2.2. Fig.
3 (c) and (d) present the SOPs for the most stable singlet and triplet
solutions with the experimental set of parameters. The triplet solution is
degenerate with the one rotated by 90${{}^{\circ}}$, so that a chiral state
can be constructed. Note that both solutions have strong angular anisotropies
(even vertical line nodes), not imposed by the $p$ or $d$ symmetries.

We have also estimated a potential effect of matrix elements in Eq. (3) by
retaining only the intra-orbital pairing, as suggested in Ref. \cite{Rice}, or
only interactions within the q1D and the q2D bands, but in either case the
singlet solution remains much more stable than the triplet one. Applying the
total susceptibility to the $\gamma$ band only results in a largely favored
singlet state, $R_{s/t}$=3.5. Magnetic anisotropy 
favors the triplet state \cite{ogata,ogata1,ogata2}. The macroscopic susceptibility tells us that the qFM
part has an easy-axis anisotropy of 20\% . For the IC part, INS finds a larger
anisotropy, of about a factor of two \cite{braden2004a}; NMR places an upper
limit at a factor of three \cite{ishida01a}. Using the latter, we find the
numbers shown in Table I. A chiral state with $\mathbf{d||z}$ would have
triplet-pair spins aligned in the $xy$ plane, and thus be disadvantaged
compared to a spin-isotropic singlet state; the ratio rises to $R_{s/t}=$7.7.
A planar state with spins perpendicular to the planes will benefit from the
easy-axis anisotropy, but not enough: the $R_{s/t}$ is still 2.9. Within
simple spin-fluctuation theory it seems almost impossible to obtain a stable
triplet solution even though the qFM signal is much sharper than previously thought.

In conclusion, we have identified the long-sought qFM fluctuations in Sr$_{2}%
$RuO$_{4}$, and, by comparing with the phonon scattering, quantitatively
determined their amplitude. Combining this qFM signal and the nesting-driven
IC response we have constructed the total magnetic susceptibility
$\chi^{\prime\prime}(\mathbf{q},E)$ at all \textbf{q}, which is consistent
with the macroscopic susceptibility, with the specific heat coefficient in the
normal state and with the $1/T_{1}T$ NMR results. Even though the
experimentally determined qFM response is stronger and sharper than thought
before, the IC component still dominates the spin-fluctuation spectrum in
Sr$_{2}$RuO$_{4}$, so that the total susceptibility   favors a singlet order
parameter for a spin-fluctuation mediated pairing. Thus, if the
superconductivity in Sr$_2$RuO$_4$ is triplet, interactions beyond simple
spin-fluctuation exchange would be required for the pairing mechanism.


This work was supported by the Deutsche Forschungsgemeinschaft through CRC
1238 Project No. B04 and by the JSPS KAKENHI No. 15H05852. I.I.M. was
supported by ONR through the NRL basic research program.


\begin{thebibliography}{10}
\bibitem{exp-det}
Experiments were performed with the $c$ direction perpendicular to the
  scattering plane in order to maximize the integration of the q2D magnetic
  signals. We used the spectrometer IN20 with focussing Heusler monochromator
  and analyzer crystals and Helmholtz coils. All measurements were performed
  with k$_{f}$=4.1${\mathring{A}}^{-1}$\ and with a pyrolytic graphite filter
  to suppress higher-order contamination.
\bibitem{noteBG}
As a result of subtracting similar count rates, the resulting error bars are
  relatively large (see Fig. 1). The BG contains an essential part that depends
  only on the scattering angle. By analyzing many $\mathbf{Q}$ scans, a BG
  function depending on the scattering angle could be fitted. The variation of
  the BG also reflects the variation of the nSF count rates indicating that
  this part of the BG arises from the finite flipping efficiency. Although in
  the analysis of the magnetic signals, the size of the magnetic scattering has
  almost exclusively been obtained from the PA, these considerations show that
  the BG is well understood and properly mastered.
\bibitem{alternativ}
The overestimation of $\chi'(0,0)$ seems to arise from the IC Lorentzian tails.
  The data can be equally well described assuming a Gaussian decay also for the
  IC amplitude, with the parameters $\chi^{\prime}_{\mathrm{IC}}$ = 214 $\mu
  _B/eV^2$, $W$=0.12 $r.l.u.$ and a constant $\Gamma_{\mathrm{IC}}$=11\ meV.
  With this description there is no contribution of IC parts to $\mathbf{q}$=0,
  but the results for the BCS eigenvalue calculation are nearly unchanged. In
  particular the singlet is still favoured with an eigenvalue (in units of
  Table I) of 78 compared to 20 for the triplet solution; $R_{s/t}$=3.9.
\end{thebibliography}
\end{document}